\documentclass[10pt,aps.prl,twocolumn,superscriptaddress,floatfix]{revtex4-1} 

\usepackage{amsmath}

\usepackage{graphics}
\usepackage{color}
\usepackage{caption}
\usepackage{subcaption}
\usepackage{float}
\usepackage{array}
\usepackage[margin=1.0in]{geometry}
\usepackage{ulem}
\usepackage{amssymb}
\usepackage{bm}
\usepackage{cancel}

\begin{document}
\title{Two-color Thermosensors based on [Y$_{1-X}$Dy$_X$(acetylacetonate)$_3$(1,10-phenanthroline) Molecular Crystals}
\author{Benjamin R. Anderson$^{*,1}$, Ray Gunawidjaja$^1$, and Hergen Eilers}
\affiliation{Applied Sciences Laboratory, Institute for Shock Physics, Washington State University,
Spokane, WA 99210-1495}
\date{\today}
\email{branderson@wsu.edu}

%

\begin{abstract}
We develop a two-color thermometry (TCT) phosphor based on [Y$_{1-x}$Dy$_x$(acetylacetonate)$_3$(1,10-phenanthroline)] ([Y$_{1-x}$Dy$_x$(acac)$_3$(phen)]) molecular crystals for use in heterogeneous materials.  We characterize the optical properties of [Y$_{1-x}$Dy$_x$(acac)$_3$(phen)] crystals at different temperatures and Dy concentrations and find that the emission is strongly quenched by increasing temperature and concentration. We also observe a broad background emission (due to the ligands) and find that  [Y$_{1-x}$Dy$_x$(acac)$_3$(phen)] photodegrades under 355 nm illumination with the photodegradation resulting in decreased luminescence intensity.  However, while decreasing the overall emission intensity, photodegradation is not found to influence the integrated intensity ratio of the ${}^4I_{15/2} \rightarrow {}^6H_{15/2}$ and ${}^4F_{9/2} \rightarrow {}^6H_{15/2}$ transitions. This ratio allows us to compute the temperature of the complex Based on the temperature dependence of these ratios we calculate that [Y$_{1-x}$Dy$_x$(acac)$_3$(phen)] has a maximum sensitivity of 1.5 \% K$^{-1}$ and our TCT system has a minimum temperature resolution of 1.8 K. Finally, we demonstrate the use of [Y$_{1-x}$Dy$_x$(acac)$_3$(phen)] as a TCT phosphor by determining a dynamic temperature profile using the emission from [Y$_{1-x}$Dy$_x$(acac)$_3$(phen)]. 

\end{abstract}

\maketitle

\vspace{1em}

\section{Introduction}
Plastic bonded explosives (PBXs) are heterogeneous materials consisting of energetic organic molecular crystals (e.g. RDX, PETN, and TNT) embedded in a polymer matrix, with possible additional additives being: plasticizers, antioxidants, taggants/markers, and friction-generating grit. These energetic organic molecular crystals can be initiated by a variety of stimuli, including thermal, mechanical or electrical. In the case of non-thermal initiation, it is believed that these stimuli first heat the energetic material, which then causes thermally-induced chemical decomposition to occur.

For mechanical shock initiation this heating is due to the shock induced formation of hot-spots within the material.  Much research has been devoted to understanding shock induced hotspot formation \cite{Chen14.01,Chen14.02,Pokharel14.01,An13.01,Wu11.01,Rae02.01,Gruzdkov01.01,Gupta95.01,Mellor93.01,Field92.01,Field92.02,Field82.01,Winter75.01,Chaudhri74.01} with the main mechanisms being: pore collapse \cite{Chaudhri74.01}, plastic deformation \cite{Pokharel14.01,Winter75.01,Field82.01}, micro-fractures, binder/crystal interface friction, and crystal/crystal interface friction. While these mechanisms have been identified to form hot-spots in shocked heterogeneous materials, it is currently unknown which of these mechanisms are responsible for chemical energy release for a given material composition and loading condition \cite{Field82.01}. 

Determining this dependence requires performing simultaneous time-resolved microstructural and thermal imaging of heterogeneous sample under shock compression. Performance of these measurements is difficult, as shock-compression events typically occur on the ns-$\mu$s time scales and microstructural imaging typically utilizes synchrotron X-Ray phase contrast imaging \cite{Luo12.01,Jensen14.01,Jensen12.01,Yeager12.01,Luo12.02}. One approach to performing in situ microstructural imaging is the recently inaugurated NNSA sponsored Dynamic Compression Sector (DCS) at the Advanced Photon Source (Argonne National Laboratory) \cite{DCSweb,DCS12.01}. The DCS provides the opportunity to perform time resolved X-ray phase contrast imaging of shock compressed heterogeneous materials, which provides the first of the two required measurements.

The second required measurement is rapid thermal imaging with ns-$\mu$s resolution. To this end we are developing temperature sensors based on two-color thermometry (TCT) phosphors, which can be used in heterogeneous materials to provide spatially resolved thermal imaging. Previously, TCT phosphors based on inorganic crystals have been used to perform temperature imaging in a wide variety of contexts  \cite{Juboori13.01,Heyes06.01,Khalid08.01,Kontis07.01,Allison97.01,Cates03.01,Gross89.01,Kontis02.01,Kontis03.01,Heyes06.01,Hasegawa07.01,Sarner05.01,Feist02.01,Feist01.01}, but for our application -- inert PBX analogues -- we desire to have organic TCT phosphors that have similar mechanical properties to the energetic materials.
 
 
The first step, towards this goal, is to develop appropriate organic molecular crystals that can be used for two-color thermometry. To achieve this goal we choose to use Dy$^{3+}$-doped yttrium ternary complexes that can be grown into molecular crystals with the first such material being [Y$_{1-x}$Dy$_x$(acetylacetonate)$_3$(1,10-phenanthroline)] ([Y$_{1-x}$Dy$_x$(acac)$_3$(phen)]). In this study we report on the optical properties of [Y$_{1-x}$Dy$_x$(acac)$_3$(phen)] at different temperatures and concentrations.

\section{Background}
\subsection{Two-Color Thermometry}
Dy$^{3+}$ is a well known lanthanide ion used for a wide variety of optical applications \cite{Kumar15.01,Gupta13.01,Jayasimhadri11.01,Ratnam10.01,Kaur14.01,Akita15.01,Juboori13.01,Heyes06.01,Khalid08.01,Kontis07.01,Allison97.01,Cates03.01,Gross89.01,Kontis02.01,Kontis03.01,Heyes06.01,Feist01.01,Anderson16.07,Gunawidjaja16.01}, with a common application being two-color thermometry (TCT) \cite{Juboori13.01,Heyes06.01,Khalid08.01,Kontis07.01,Allison97.01,Cates03.01,Gross89.01,Kontis02.01,Kontis03.01,Heyes06.01,Feist01.01}. TCT using Dy relies on two closely spaced Dy energy levels (${}^4I_{15/2}$ and ${}^4F_{9/2}$) to act as a probe of temperature. When a Dy doped material is optically pumped (with an appropriate wavelength) the ions are excited into the ${}^4F_{9/2}$ energy level, from which they can also populate the ${}^4I_{15/2}$ energy level due to thermal effects. The ratio of the populations in the ${}^4I_{15/2}$ energy level $n_2$,  and the population in the ${}^4F_{9/2}$ energy level $n_1$, is determined by the Boltzmann distribution, such that their ratio is:

\begin{align}
\frac{n_2}{n_1}=e^{-\Delta E/k_BT} \label{eqn:dist}
\end{align}
where $\Delta E$ is the energy difference between the ${}^4I_{15/2}$ and ${}^4F_{9/2}$ energy levels (approximately 0.115 eV for aqua ions \cite{Carnall68.01,Carnall79.01,Bunzli10.01,Bunzli10.02}), $k_B$ is Boltzmann's constant, and $T$ is the temperature.

This ratio between the ${}^4I_{15/2}$ and ${}^4F_{9/2}$ energy levels can be determined using emission spectroscopy; as the ratio of the ${}^4I_{15/2} \rightarrow {}^6H_{15/2}$ and ${}^4F_{9/2} \rightarrow {}^6H_{15/2}$ transition is proportional to the ratio of populations:

\begin{align}
\frac{I_2}{I_1}&\propto \frac{n_2}{n_1} \nonumber
\\ &= Ae^{-\Delta E/k_BT}, \label{eqn:Irat}
\end{align}
where $A$ is a proportionality constant.

\subsection{Ligand Enhanced Luminescence}
While Dy$^{3+}$ has advantageous optical properties for TCT, on its own Dy$^{3+}$ (and the other lanthanide ions) has a small molar absorption coefficient, making it an inefficient phosphor. To overcome this limitation much work has been done on embedding different lanthanide ions in ligands such that the ligands (which have large absorption coefficients) absorb the pump light and subsequently transfer the absorbed energy to the lanthanide ion \cite{Sager65.01,Dexter53.01,Xin04.01,Klink00.01,Latva97.01,Feng09.01,Bunzli10.01,Adronov00.01,Dickens98.01,Bunzli89.01,Shavaleev10.01,Eliseeva10.01,Puntus09.01,Bunzli14.01,Bunzli16.01,Bunzli15.01}. This energy transfer is typically radiationless and can occur via several possible mechanisms: electron exchange \cite{Dexter53.01,Adronov00.01}, dipole-dipole interactions (both allowed \cite{Forster65.01,Forster59.01,Forster48.01,Forster46.01} and unallowed transitions \cite{Dexter53.01,Dexter54.01}), dipole-quadrapole interactions \cite{Dexter53.01}, and excitons \cite{Dexter51.01,Heller51.01}.    

\begin{figure}
 \centering
 \includegraphics{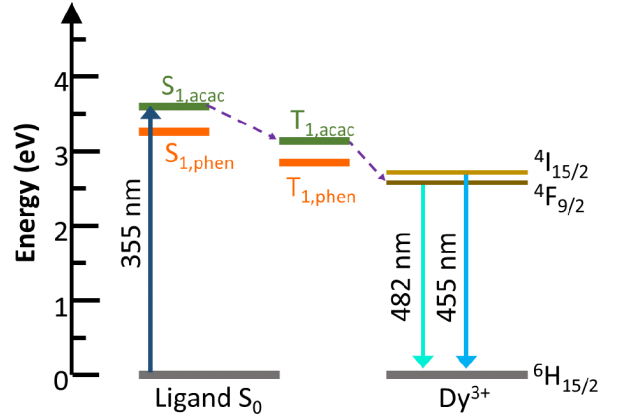}
 \caption{Energy level schematic of [Dy(acac)$_3$(phen)] displaying energy transfer from the ligands' triplet state to Dy$^{3+}$'s lowest excited state.}
 \label{fig:Elevels}
\end{figure}

This energy transfer has been empirically found to be most efficient if the ligand's lowest triplet energy level is between 0.23 eV and 0.62 eV greater than the lanthanide's primary luminescent state, as energy back-transfer is more likely to occur for energy differences of $< 0.23$ eV \cite{Sato70.01,Latva97.01}. One such material, satisfying this criterion is [Dy(acetylacetonate)$_3$(1,10-phenanthroline)] ([Dy(acac)$_3$(phen)]) \cite{Feng09.01,Tang15.01,Chen11.02,Choy07.01,Brito09.01}, which is the subject material of this study. To demonstrate the energy level spacing of the ligands and Dy ion we provide a simplified energy level diagram in Figure \ref{fig:Elevels} where the energy levels are taken from the literature \cite{Peng05.01,Sager65.01,Xin06.01,Feng09.01,Bunzli10.01}. Note that in general there are many more energy levels for the ligands and Dy than displayed in Figure \ref{fig:Elevels}, but we simplify the diagram to highlight the relevant transitions.

From a review of the literature \cite{Peng05.01,Sager65.01,Xin06.01,Feng09.01,Bunzli10.01} we find that the acac ligand has an $S_0\rightarrow S_1$ transition corresponding to $\approx 3.596$ eV (345 nm), phen has an $S_0\rightarrow S_1$ transition corresponding to $\approx 3.263$ eV (380 nm), and Dy$^{3+}$'s primary luminescent state (${}^4F_{9/2}$) is at $\approx 2.583$ eV (480 nm). In our study we use a frequency tripled Nd:YAG laser with a wavelength of 355 nm for our excitation source, which places the excitation wavelength nearly resonant with acac's $S_0\rightarrow S_1$ transition. This means that during excitation the primary absorption is due to the acac ligand, which can then nonradiatively transition to the lowest triplet state $T_1$ (energy $\approx 3.134$ eV) via intersystem crossing. Once in the triplet state the energy can be transferred from the ligand to the Dy$^{3+}$ ion with the approximate energy difference between acac's $T_1$ state and the ${}^4F_{9/2}$ state of Dy$^{3+}$ being 0.554 eV. Since this energy difference is greater than 0.23 eV, backwards energy transfer will be minimal.

\section{Method}

\subsection{Materials}
The primary component of our two-color thermosensors is [Dy(acetylacetonate)$_3$(1,10-phenanthroline)] ([Dy(acac)$_3$(phen)]). [Dy(acac)$_3$(phen)] has previously been investigated for use as a single molecule magnet \cite{Tang15.01,Chen11.02}, a phosphor for white OLEDs \cite{Choy07.01,Brito09.01}, and a NIR optical phosphor for use in telecommunications \cite{Feng09.01}. In these previous studies the authors prepared the material with a Dy concentration of 100 mol\%, while in our study we dilute [Dy(acac)$_3$(phen)] with [Y(acac)$_3$(phen)]. We perform this dilution in order to characterize the influence of Dy concentration on the material's optical properties. This combination of [Dy(acac)$_3$(phen)] and [Y(acac)$_3$(phen)] results in a hybrid molecular crystal, which we denote [Y$_{1-x}$Dy$_x$(acac)$_3$(phen)]. 

To prepare [Y$_{1-x}$Dy$_x$(acac)$_3$(phen)] we use the following procedure \cite{Chen11.02}: First we prepare a solution of acetylacetone (0.06 M) and 1,10-phenanthroline (0.02 M) in a DMF/methanol mixture (1:1 vol./vol. ratio). To this solution we add an aqueous solution of potassium tert-butoxide (KOtBu) (0.06 M) at an equal volume ratio. Once mixed we add another equal volume aqueous solution containing Dy(NO$_3$)$_3$.5H$_2$O and Y(NO$_3$)$_3$.6H$_2$O in different Dy/Y mole ratios such that the total concentration is 0.02 M. For this study we make [Y$_{1-x}$Dy$_x$(acac)$_3$(phen)] with Dy concentrations of 1 mol\%, 5 mol\%, 10 mol\% and 15 mol\%. Gradually white precipitates appear and the suspension is allowed to age under stirring at room temperature for 4-6 hours.  The precipitates are isolated via centrifugation (6000 rpm for 3 min) and finally dried in vacuum at 80 $^\circ$C for $\approx 12$ hours. The resulting powder consists of [Y$_{1-x}$Dy$_x$(acac)$_3$(phen)] molecular crystals (MCs).

To characterize the structure of the [Y$_{1-x}$Dy$_x$(acac)$_3$(phen)] MCs we use X-ray diffraction (XRD) with a Cu-K$\alpha$ radiation source ($\lambda$=1.5418 \AA) operated at 40 kV and 40 mA. The X-ray beam is monochromated using an X'Celerator monochromator (PANalytical B.V.) and collimated using a fixed divergence slit with 0.04 radian Soller slits, 0.5$^{\circ}$ divergence slit and 10 mm mask. The diffracted X-rays are detected using a PIXcel3D detector (PANalytical B.V.) with the resulting diffraction patterns being analyzed using X'Pert HighScore Plus, Version 3.0 (PANalytical B.V.). Additional characterization is performed by optical microscopy of the MC powder.

\begin{figure}
 \centering
 \includegraphics{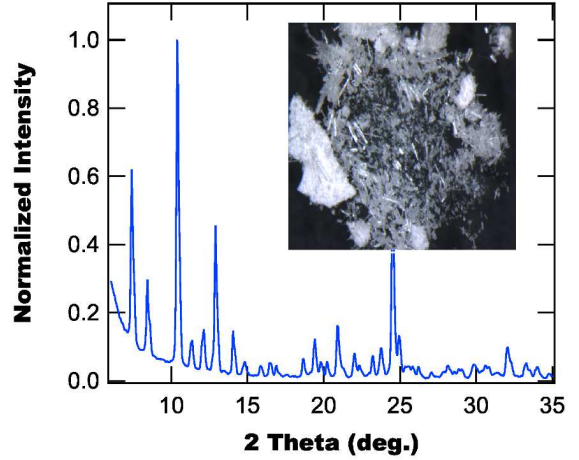}
 \caption{XRD pattern for [Dy(acac)$_3$(phen)] with multiple sharp peaks visible. Inset: Optical microscope image of crystals.}
 \label{fig:XRD}
\end{figure}

Figure \ref{fig:XRD} shows a representative XRD pattern for [Dy(acac)$_3$(phen)], with the pattern displaying sharp peaks consistent with a polycrystalline structure. This structure is also confirmed by optical microscopy, with the inset of Figure \ref{fig:XRD} being an optical image displaying multiple crystals of varying shapes and sizes. To determine the crystal parameters (structure, space group, lattice size, etc.) of the material we analyze the XRD pattern for each Dy concentration and find that the crystal parameters are within uncertainty of each other for all Dy concentrations. We therefore average the parameters and tabulate their average value in Table \ref{tab:param}. These parameters are found to be consistent with previous XRD measurements of [Dy(acac)$_3$(phen)] MCs \cite{Chen11.02}.

\begin{table}
\caption{Average crystal parameters for [Y$_{1-x}$Dy$_x$(acac)$_3$(phen)] determined from XRD. These parameters are found to be invariant with Dy concentration (within experimental uncertainty). }
\label{tab:param}
\centering
 \begin{tabular}{cc}
 \hline 
\textbf{Parameter} & \textbf{Value} \\ \hline
 Crystal System & Monoclinic \\
  Space Group & P2$_1$/c\\
 $a$ (\AA) & 16.1 \\
 $b$ (\AA) & 21 \\
 $c$ (\AA) & 9.4 \\
 $\alpha$ & 90$^{\circ}$ \\
 $\beta$ & 116$^{\circ}$ \\
 $\gamma$ & 90$^{\circ}$ \\
 V (\AA$^3$) & 2841.82\\ \hline
 \end{tabular}

\end{table}

\subsection{Spectroscopy}
\begin{figure*}
 \centering
 \includegraphics{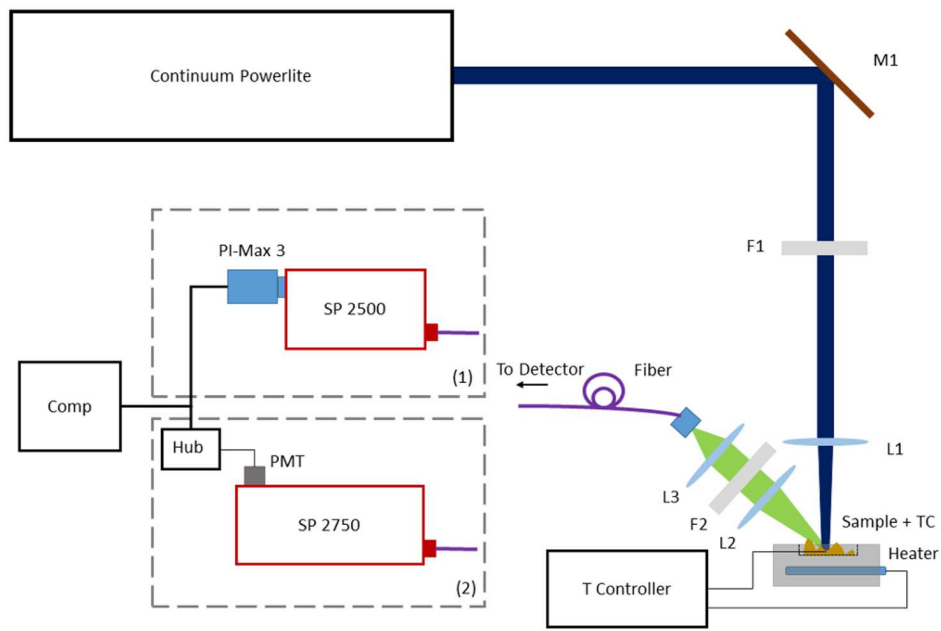}
 \caption{Schematic of experimental setup with both the gated (1) and ungated (2) spectrometers. F1: 355 nm Laser line filter, F2: long-pass filter (400 nm cut-on), L1,L2,L3: Lenses, M1: mirror, Hub: Acton Spectra Hub.}
 \label{fig:setup}
\end{figure*}

To measure the luminescence spectra of [Y$_{1-x}$Dy$_x$(acac)$_3$(phen)] at different temperatures we use a custom fluorescence spectroscopy system and powder heater. The spectroscopy system consists of a frequency-tripled Nd:YAG laser (Continuum Powerlite II 8000, 355 nm, 8 ns), focusing and collection optics, and two different spectrometers (one for temporally gated spectral measurements and the other for ungated measurements). The spectrometer for ungated measurements consists of an Acton SpectraPro 2750i monochromator, attached current PMT, and a Spectrahub interface, while the spectrometer for gated measurements consists of a Princeton Instruments PI-Max 3 ICCD connected to an Acton SpectraPro 2500i spectrometer. In addition to spectral measurements the ungated system is also used to measure florescence lifetimes with the PMT output  connected to a Tektronix DPO 4104 oscilloscope. Figure \ref{fig:setup} shows a schematic of the spectroscopy system.

In order to evenly heat our [Y$_{1-x}$Dy$_x$(acac)$_3$(phen)] molecular crystal powder we use a custom powder heater. The heater consists of a block of Aluminum (2" D $\times$ 1" H)  containing a 1 cm diameter indentation in which the powder is placed.  This configuration allows for heating the powder from both the bottom and sides. For heating we use a 120 W canister heater which is controlled by an Omega CN32PT-220 PID controller with a K-type thermocouple providing temperature feedback.

\section{Results and Discussion}

\subsection{Ungated Spectra}
\begin{figure}
\centering
\includegraphics{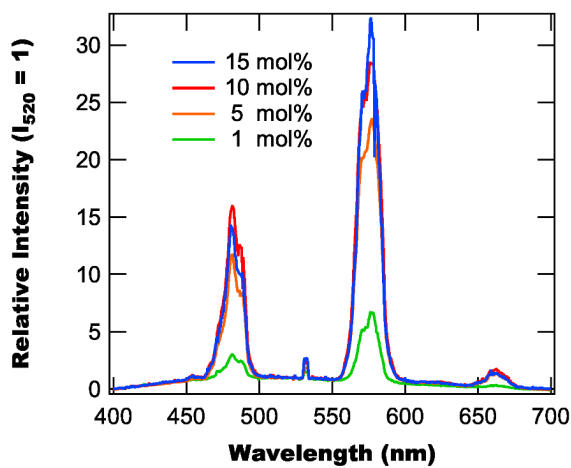}
\caption{Scaled emission spectra of [Y$_{1-x}$Dy$_x$(acac)$_3$(phen)] MCs for different Dy concentrations. The spectra consist of four broad peaks from the Dy ions superimposed on a broad background emission from the organic ligand. The small peak at 532 nm is residual second harmonic from the pump laser.}
\label{fig:ungated}
\end{figure}

We begin our study of [Dy(acac)$_3$(phen)] by considering its ungated emission spectra at different Dy concentrations as shown in Figure \ref{fig:ungated}. Note that the intensity in Figure \ref{fig:ungated} is scaled such that the intensity at 520 nm is normalized.  From Figure \ref{fig:ungated} we find that the emission spectra consist of four broad peaks corresponding to emission from Dy with the peak centers being 453 nm (${}^4I_{15/2} \rightarrow {}^6H_{15/2}$), 482 nm (${}^4F_{9/2} \rightarrow {}^6H_{15/2}$), 577 nm (${}^4F_{9/2}\rightarrow {}^6H_{13/2}$), and 660 nm (${}^4F_{9/2} \rightarrow {}^6H_{11/2}$ ).

\begin{figure}
 \centering
 \includegraphics{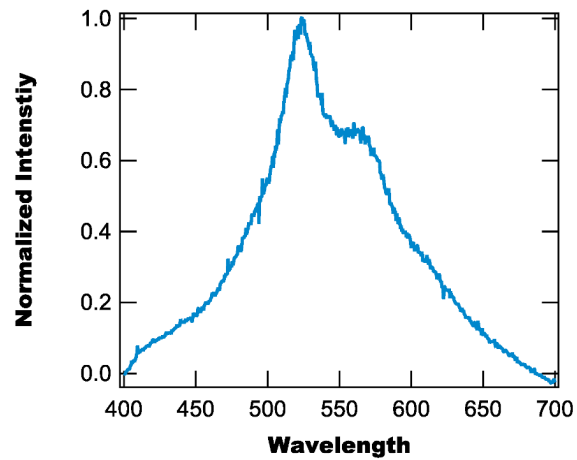}
 \caption{Normalized emission spectrum from a [Y(acac)$_3$(phen)] sample with 355 nm excitation. The emission is broad and spans the whole spectral region of interest.}
 \label{fig:lig}
\end{figure}

In addition to the four Dy peaks in Figure \ref{fig:ungated}, we also observe a broad background emission centered at 520 nm (which is found to originate from the organic ligand) and a peak at 532 nm (corresponding to residual second harmonic in the pump). The ligand emission is isolated from the Dy emission by measuring a pure [Y(acac)$_3$(phen)] sample, as shown in Figure \ref{fig:lig}.

\subsection{Gated Spectra}
Given the proposed application of [Y$_{1-x}$Dy$_x$(acac)$_3$(phen)] as a TCT phosphor it is necessary to remove the ligand's emission from the measured spectrum. This removal allows for an accurate probe of the Dy populations in the ${}^4I_{15/2}$ and ${}^4F_{9/2}$ energy levels. While the ligand's emission can easily be subtracted using multi-peak fitting when performing spectral measurements, this technique is not an option when performing time-resolved TCT imaging. Instead, for time-resolved TCT imaging, we rely on the different luminescence lifetimes of the Dy$^{3+}$ ions and the ligand. 

\begin{figure}
\centering
\includegraphics{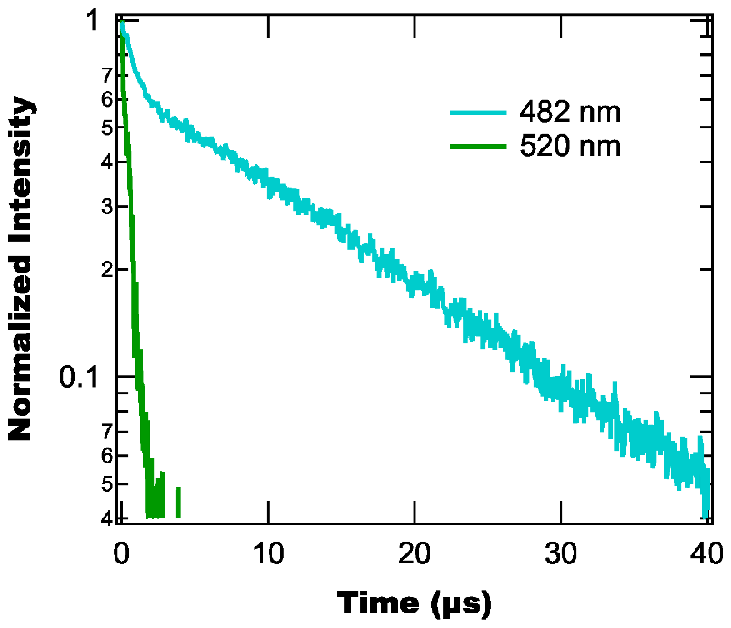}
\caption{Normalized 482 nm and 520 nm emission intensity from a [Y$_{0.99}$Dy$_{0.01}$(acac)$_3$(phen)] sample as a function of time. The emission at 482 nm is found to be bi-exponential with lifetimes of 16 $\mu$s and 0.6 $\mu$s, while the 520 nm emission is found to follow a single exponential with a lifetime of 0.6 $\mu$s. The bi-exponential behavior of the 482 nm intensity is due to emission from both the lanthanide and ligands, while the 520 nm intensity follows a single exponential as it consists of emission from the ligands only.}
\label{fig:gtime}
\end{figure}

To demonstrate the different lifetimes we plot the normalized emission intensity (for a 1 mol\% sample) at 482 nm (primarily due to Dy) and 520 nm (due to ligand) as a function of time in Figure \ref{fig:gtime}. From Figure \ref{fig:gtime} we find that the 520 nm intensity follows a single exponential -- with lifetime 0.6 $\mu$s --  while the 482 nm intensity is bi-exponential -- with lifetimes of 16 $\mu$s and 0.6 $\mu$s. Given the value of the short lifetime (for the 482 nm intensity) and the observation that [Y(acac)$_3$(phen)] has emission at 482 nm, we conclude that the fast component of the bi-exponential is due to the ligand, while the long lifetime component is due Dy's 
${}^4F_{9/2} \rightarrow {}^6H_{15/2}$ transition. The order of magnitude difference in emission lifetime between the ligands and the Dy ion means that we can remove the ligands' emission by using gated measurements with a delay greater than the emission lifetime of the ligands.

\begin{figure}
 \centering
 \includegraphics{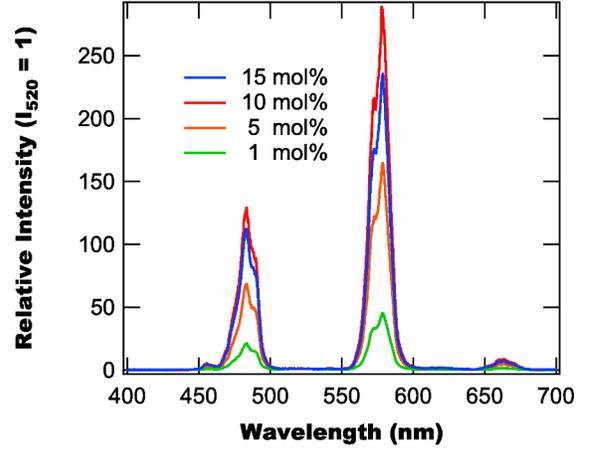}
 \caption{Scaled gated emission spectra of [Y$_{1-x}$Dy$_x$(acac)$_3$(phen)] MCs for different Dy concentrations}
 \label{fig:gatespec}
\end{figure}

Using a gate delay of 1 $\mu$s we re-perform the spectral measurements on the different [Y$_{1-x}$Dy$_x$(acac)$_3$(phen)] samples (scaled spectra are shown in Figure \ref{fig:gatespec}) and obtain improved contrast between the Dy and ligand emission as the ligand's emission is mostly removed. From Figure \ref{fig:gatespec} we find that the contrast between the emission at 520 nm and 482 nm is approximately 120, which is an order of magnitude larger than for the ungated spectral measurements (Figure \ref{fig:ungated}). This increased contrast minimizes the influence of the ligand's emission and allows for more accurate determination of the intensity ratios used in Equation \ref{eqn:Irat}.

\subsection{Concentration Dependence}
\begin{figure}
\centering
\includegraphics{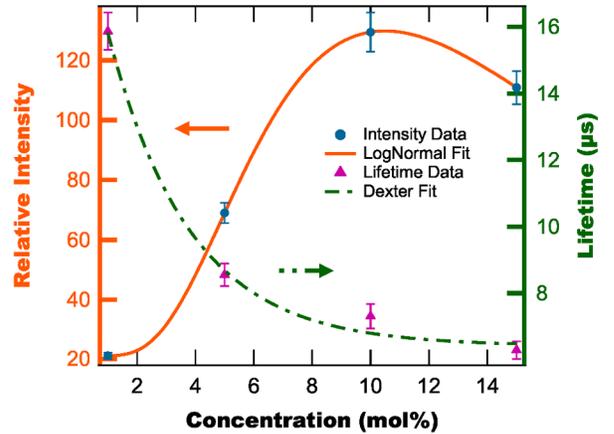}
\caption{Relative intensity and lifetime of the emission at 482 nm as a function of Dy concentration. The relative intensity is fit to a lognormal function and the lifetime is fit to Equation \ref{eqn:dex}.}
\label{fig:CQ}
\end{figure}

From Figure \ref{fig:gatespec} we observe that, while the relative intensity of the Dy emission increases as the concentration increases from 1 mol\% to 10 mol\%, it begins to decrease as the concentration is further increased.  This behavior arises as the increase in concentration initially is beneficial (providing more ions for luminescence), but as the concentration is further increased quenching becomes dominate, and limits the emission from the material.  Figure \ref{fig:CQ} demonstrates this effect in the relative intensity at 482 nm as a function of concentration, with a fit to a lognormal function as a guide for the eye. From Figure \ref{fig:CQ} we find that the 10 mol\% sample provides the most emission intensity.

\begin{figure*}
 \begin{subfigure}[t]{0.47\textwidth}
 \centering
\includegraphics{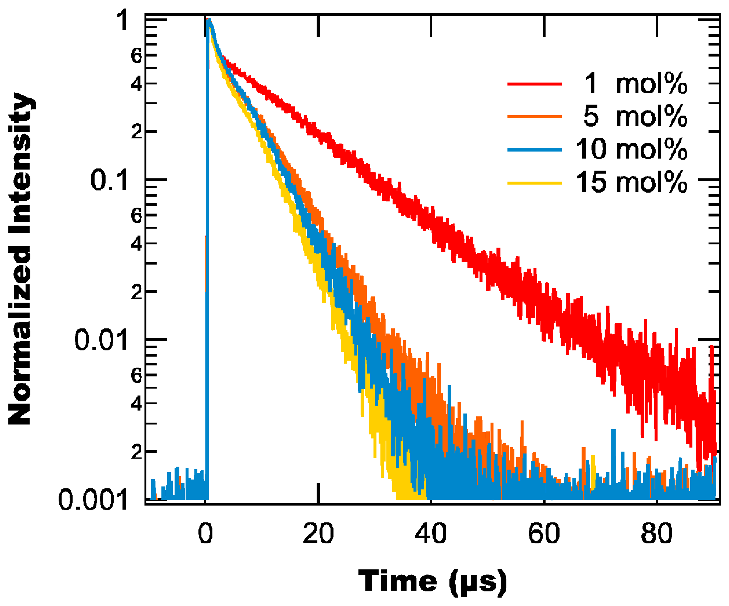}
\caption[a]{}
\label{fig:482}
\end{subfigure}
 \begin{subfigure}[t]{0.47\textwidth}
  \centering
\includegraphics{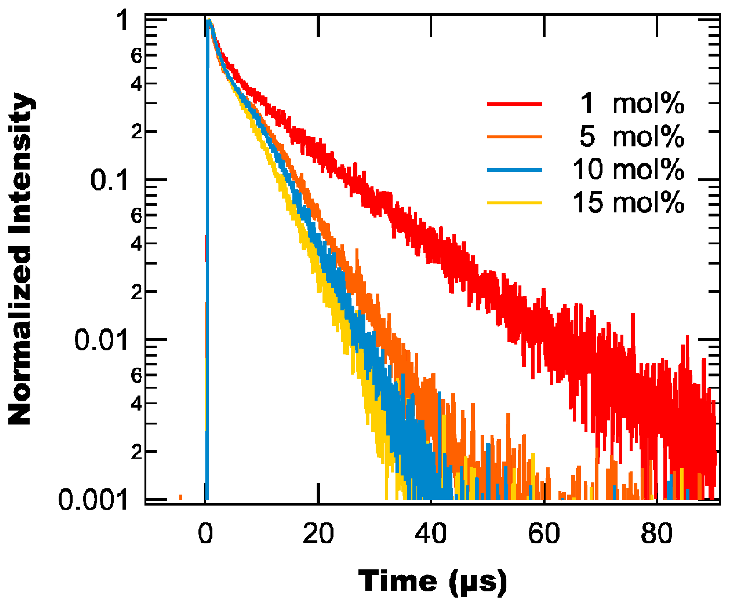}
\caption[b]{}
\label{fig:575}
\end{subfigure}
\caption{Normalized luminescence intensity as a function of time for different Dy concentrations at wavelengths of 482 nm (a) and 575 nm (b). Note that the fast decay seen at early times has a concentration independent lifetime of $\approx 0.6$ $\mu$s, which is consistent with emission from [Y(acac)$_3$(phen)].}
\end{figure*}

To further characterize concentration quenching in [Y$_{1-x}$Dy$_x$(acac)$_3$(phen)], we measure the luminescence intensity as a function of time after excitation to determine the luminescence lifetime at different Dy concentrations. Figures \ref{fig:482} and \ref{fig:575} show the emission intensity as a function of time after excitation for the different concentrations. From both Figures \ref{fig:482} and \ref{fig:575} we find that the luminescence decays as a bi-exponential function with the fast decay occurring with a concentration independent lifetime of $\approx 0.6$ $\mu$s and the slower decay having a concentration dependent lifetime, which decreases with increasing concentration. The observation of the long lifetime depending on the Dy concentration and the short lifetime being invariant with concentration is additional evidence that the short lifetime component is due to the ligands, while the long lifetime component is due to the Dy ion.

Using simple exponential fits to the data in Figures \ref{fig:482} and \ref{fig:575} we determine the lifetime as a function of Dy concentration, which is shown in Figure \ref{fig:CQ}. From Figure \ref{fig:CQ} we find that as the concentration increases the lifetime becomes shorter, with the largest change occurring when going from 1 mol\% to 5 mol\%.

The influence of concentration quenching on lanthanide luminescence lifetimes has been studied in depth  \cite{Kharabe15.01,Dexter53.01,Dexter54.01,Joos12.01,Meza14.01,Ju13.01,lakow06.01,Forster65.01,Forster59.01,Forster48.01,Forster46.01} with the two main mechanisms of concentration quenching in solid-state materials being: long-range Energy Transfer (LRET) \cite{lakow06.01,Forster65.01,Forster59.01,Forster48.01,Forster46.01}, and short-range Dexter electron exchange \cite{Dexter54.01}. These mechanisms result in different functional behavior of the lifetime as a function of concentration. Assuming that the ion has an unquenched lifetime of $\tau_0$, the lifetime $\tau$ (as a function of concentration $C$) is:
\begin{align}
\tau=\frac{\tau_0}{1+\left(\frac{C}{C_H}\right)^\alpha}, \label{eqn:ET}
\end{align}
for LRET with $C_H$ being the half-quenching concentration and $\alpha$ being an exponent that depends on the energy transfer mechanism (e.g. dipole-dipole interactions, quadrapole-dipole interactions). For Dexter electron exchange the lifetime as a function of concentration is given by:
\begin{align}
\tau=\frac{\tau_0}{1+k_De^{-(C/C_0)^{-1/3}}}, \label{eqn:dex}
\end{align}
where $c_0$ is the critical concentration \cite{Bierwagen16.01} and $k_D$ is a material dependent constant. From the functional forms of Equations \ref{eqn:ET} and \ref{eqn:dex} we note that LRET results in the lifetime asymptotically approaching zero, while the Dexter mechanism has the lifetime asymptotically approaching a nonzero value. Given these differences in functionality of Equations \ref{eqn:ET} and \ref{eqn:dex}, and the observed behavior of the lifetime in Figure \ref{fig:CQ}, we conclude that the Dexter mechanism's concentration dependence best matches the observed experimental data. Therefore we fit the lifetime data in Figure \ref{fig:CQ} to Equation \ref{eqn:dex} and observe good agreement.

The result of the experimental lifetimes following Equation \ref{eqn:dex} is surprising, as typically the other quenching mechanisms are more dominant \cite{lakow06.01} as they are longer range interactions than Dexter electron exchange. Typically Dexter electron exchange based quenching doesn't occur until the average spacing of ions is on the order of 10's of angstroms, as the Dexter mechanism is proportional to the overlap of the acceptor and donor wavefunctions \cite{Dexter54.01}. 

Given this spatial dependence it is necessary to estimate the ion spacing in order to determine if Dexter electron exchange is a valid hypothesis. To do so we first estimate the particle density using the lattice volume (see Table \ref{tab:param}), which gives a particle density of $n_p=3.52\times10^{26}$ m$^{-3}$. This density represents the number of molecular complexes in a given volume, but not the density of Dy ions. To get the density of Dy ions we need to scale the particle density by the doping percentage, which gives the ion density to be $n_i=cn_p$, where $c$ is the doping percentage. While a precise calculation of the mean inter-ion distance $\langle r\rangle$, from the concentration is difficult, we can get a rough estimate using the formula $\langle r\rangle \sim 1/n_i^{1/3}$. Using this formula we get an inter-ion distance of 65.7 \AA{} for the 1 mol\% sample and 26.7 \AA{} for the 15 mol\% sample. These spacings suggest that the Dexter mechanism is a valid hypothesis for our observed concentration dependence. However, further study (using more concentrations and a calculation of the exchange radius) is required to verify this result.

\subsection{Temperature Dependence}
\subsubsection{Emission Spectra}
\begin{figure}
 \centering
 \includegraphics{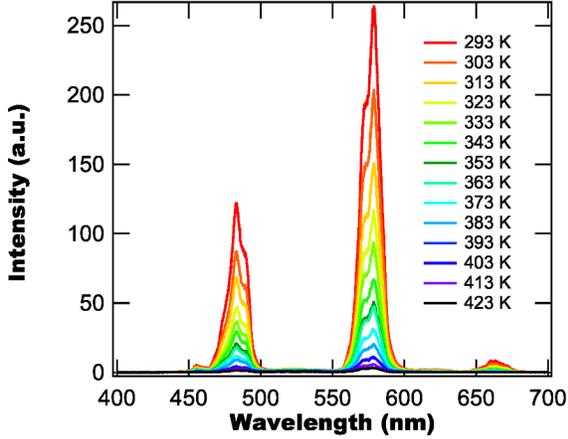}
 \caption{Emission spectra of a [Y$_{1-x}$Dy$_x$(acac)$_3$(phen)] (Dy concentration of 10 mol\%) at 14 different temperatures. The emission displays strong thermal quenching.}
 \label{fig:specT}
\end{figure}

With the concentration dependence of the emission of [Y$_{1-x}$Dy$_x$(acac)$_3$(phen)] characterized, we now turn to consider the influence of temperature on its emission. To do so we use the gated spectroscopy system to measure the temperature dependent emission of [Y$_{1-x}$Dy$_x$(acac)$_3$(phen)] at 14 different temperatures ranging from 293 K to 423 K. Figure \ref{fig:specT} shows an example set of emission spectra for the 10 mol\% sample. From Figure \ref{fig:specT} we find that the emission spectra display strong thermal quenching, which is seen in the emission from all concentrations tested.

\begin{figure}
\centering
\includegraphics{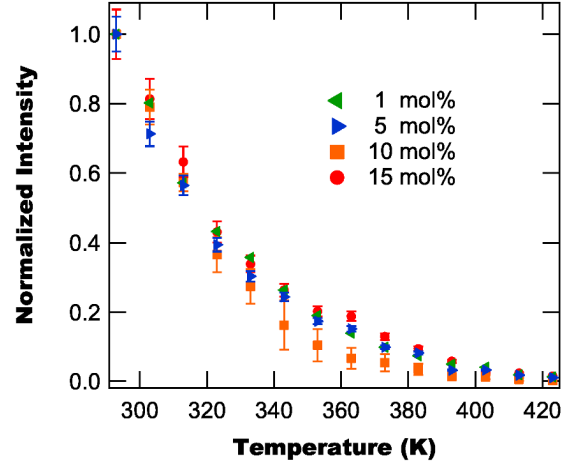}
\caption{Normalized emission intensity at 482 nm as a function of temperature for different Dy concentrations, with the intensity normalized to the room temperature value. The emission shows strong thermal quenching. }
\label{fig:TI}
\end{figure}

We characterize the material's thermal quenching by plotting the emission intensity at 482 nm as a function of temperature for all four concentrations in Figure \ref{fig:TI}, with the intensity normalized such that the room temperature intensity is unity. From Figure \ref{fig:TI} we find that the intensity as a function of temperature decays exponentially with each concentration's curve being within uncertainty of each other. Fitting each curve to a simple exponential and taking the weighted average of the decay constants over all four concentrations we find that the average decay constant is $36.89 \pm 0.84$ K. 

To consider what this temperature dependence means practically, we assume that we have a detection system capable of accurately measuring down to 1\% of the room temperature signal. Using the decay constant determined above, this dynamic range gives a functional maximum temperature of approximately 463 K. However, for our current detection system we find that for temperatures above 423 K the signal to noise ratio is unacceptably small ($\approx 2$ for the 455 nm peak and $\approx 16$ for the 482 nm peak). Thus our current maximum temperature limit is $\approx 423$ K.

\subsubsection{Luminescence Lifetime}

\begin{figure}
\centering
\includegraphics{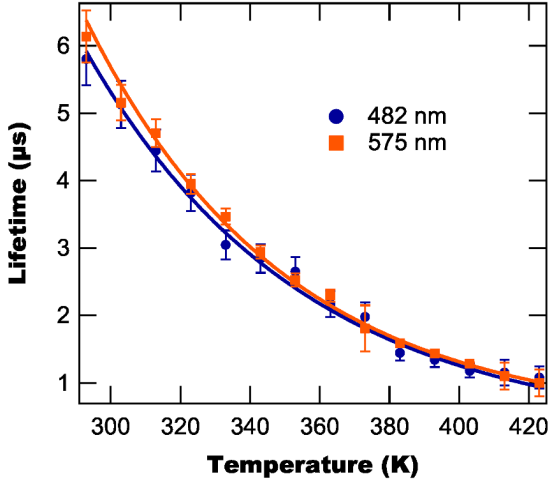}
\caption{Lifetimes of the luminescence emission at 482 nm and 575 nm as a function of temperature for [Y$_{0.
9}$Dy$_{0.1}$(acac)$_3$(phen)]. As the temperature increases the lifetimes decrease due to thermal quenching.}
\label{fig:LTTemp}
\end{figure}

In addition to the thermal quenching affecting the emission intensity we also find that it affects emission lifetime, which is consistent with previous measurements of other lanthanide-doped complexes \cite{Ueda15.01,Bierwagen16.01,Joos12.01,Yu14.01}.  We characterize the influence of thermal quenching on the lifetime of  [Y$_{1-x}$Dy$_x$(acac)$_3$(phen)] by measuring the lifetime of the emission at 482 nm and 575 nm for the 10 mol\% sample, as shown in Figure \ref{fig:LTTemp}. From Figure \ref{fig:LTTemp} we find that thermal quenching in [Y$_{0.9}$Dy$_{0.1}$(acac)$_3$(phen)] results in the lifetime decreasing with increasing temperature.

This decrease in lifetime with increasing temperature is typically modeled as a barrier process with the rate of energy loss from the excited state $k$, being 

\begin{align}
 k&=k_0+k_{NR}(T), \nonumber
 \\ &=k_0+k_1e^{-\Delta E_T/kT}, \label{eqn:kT}
\end{align}
where $k_0$ is the excited state's natural decay rate, $k_{NR}(T)=k_1e^{-\Delta E_T/kT}$ is the rate of nonradiative energy transfer as a function of temperature $T$, with $\Delta E_T$ being the energy transfer barrier and $k_1$ being a rate related to non-radiative energy transfer from the excited state. Recalling that the lifetime of the excited state is given by  $\tau=1/k$, we use Equation \ref{eqn:kT} to give the lifetime as a function of temperature:

\begin{align}
 \tau&=\frac{1}{k_0+k_1e^{-\Delta E_T/kT}}, \nonumber
 \\ &=\frac{\tau_0}{1+\frac{\tau_0}{\tau_1}e^{-\Delta E_T/kT}},\label{eqn:LTTQ}
\end{align}
where $\tau_i=1/k_i$.

Fitting the lifetime data in Figure \ref{fig:LTTemp} to Equation \ref{eqn:LTTQ} we determine the different parameters for both the 482 nm and 575 nm peaks with the results tabulated in Table \ref{tab:TQ}. From Table \ref{tab:TQ} we find that the parameters for both the 482 nm and 575 nm peak are within uncertainty of each other. This is expected as both peaks correspond to transitions from the ${}^4F_{9/2}$ energy level.

\begin{table}
\caption{Fit parameters determined from lifetimes as a function of temperature. The parameters for the lifetime at 482 nm and 575 nm are found to be within uncertainty of each other.}
\label{tab:TQ}
\centering
 \begin{tabular}{ccc}
 \hline
  \textbf{Parameter} & \textbf{482 nm}  & \textbf{575 nm} \\ \hline
  $\tau_0$ ($\mu$s)  & $11.6\pm4.5$  &  $13.8 \pm3.9$ \\
  $\tau_1$(ns)   & $4.2\pm2.5$  & $5.3\pm1.7$ \\
  $\Delta E_T$ (10$^{-3}$ eV) & $200\pm 28$ &  $193 \pm 13$ \\
  \hline
 \end{tabular}
\end{table}

\subsubsection{Intensity Ratios} \label{sec:rat}
At this point we have characterized the emission properties of [Y$_{1-x}$Dy$_x$(acac)$_3$(phen)] for different temperatures and concentrations. The next step towards using [Y$_{1-x}$Dy$_x$(acac)$_3$(phen)] as a TCT phosphor is to characterize the excited state ratios between the ${}^4I_{15/2}$ and ${}^4F_{9/2}$ states as a function of temperature. This ratio is probed by measuring the ratio between the emission peak at 455 nm (${}^4I_{15/2} \rightarrow {}^6H_{15/2}$) and 482 nm (${}^4F_{9/2} \rightarrow {}^6H_{15/2}$). In our study -- to improve the signal to noise ratio -- we use the spectrally integrated peak intensities to calculate the peak ratios with the 455 nm peak having an integration range of 450 nm to 460 nm and the 482 peak having an integration range of 465 nm to 500 nm.

\begin{figure}
\centering
\includegraphics{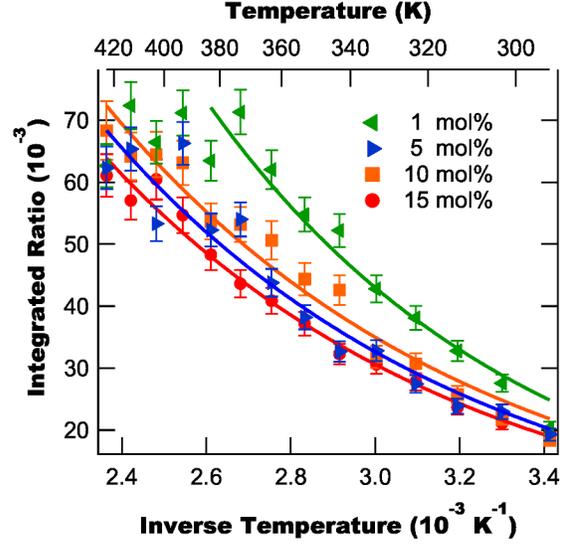}
\caption{Integrated peak ratio (455 nm/482 nm) as a function of temperature for four different Dy concentrations}
\label{fig:TR}
\end{figure}

Using the gated spectra measured at different temperatures for each concentration we integrate the spectral peaks and calculate their ratios.  The ratios of the integrated peaks are shown in Figure \ref{fig:TR} as a function of inverse temperature. Note that for the 1:99 sample the intensity at temperatures above 383 K becomes so weak as to make the ratios unreliable. These ratios are therefore not used when performing fitting.

From Figure \ref{fig:TR} we find that for all four concentrations the integrated ratio behaves as an exponential with inverse temperature, which is consistent with Equation \ref{eqn:Irat}. Fitting the ratios in Figure \ref{fig:TR} to Equation \ref{eqn:Irat} we determine both the amplitude parameter $A$ and the energy barrier $\Delta E$, which are tabulated in Table \ref{tab:fit}.

\begin{table}
\caption{Exponential fit parameters determined from Figure \ref{fig:TR} for different Dy concentrations}
\label{tab:fit}
\centering
 \begin{tabular}{ccc}
 \hline
  \textbf{Dy Concentration (mol\%)}  &  $A$  & $\Delta E$ ($10^{-3}$ eV) \\ \hline
  1 & 2.30 $\pm$ 0.87  &  114 $\pm$ 14 \\
  5 & 1.06 $\pm$ 0.31  &  100.1 $\pm$ 9.4 \\
  10  & 1.08 $\pm$ 0.20  &  98.4 $\pm$ 6.2 \\
  15  & 0.98 $\pm$ 0.13  &  99.7 $\pm$ 4.3 \\ \hline
 \end{tabular}
\end{table}

From Table \ref{tab:fit} we find that the energy differences are all within uncertainty of each other and have a weighted mean of $100.2(\pm 3.2) \times 10^{-3} $ eV. This energy difference is significantly less than the expected energy difference of $\approx 0.15$ eV (energy difference between photons with wavelengths of 482 nm and 455 nm). While the measured difference is less than the difference expected due to the peak wavelengths, it is actually consistent with the energy difference determined for Dy$^{3+}$ in a water solution: $\Delta E \approx 0.1147$ eV (energy difference between photons with wavelengths of 475 nm and 455 nm) \cite{Carnall68.01,Carnall79.01,Bunzli10.01,Bunzli10.02}. 

To understand why the energy difference measured in this study (and seen in the literature \cite{Carnall68.01,Carnall79.01,Bunzli10.01,Bunzli10.02,Gross89.01,Khalid08.01,Feist00.01}) differs from the the value determined using the peak wavelengths, we note that the peak spacing (455 nm vs. 482 nm) does not directly represent the spacing between the ${}^4I_{15/2}$ and ${}^4F_{9/2}$ energy levels. In reality each broad spectral peak consists of multiple overlapping smaller peaks that arise due to transitions from the excited state into the Stark-split ${}^6H_{15/2}$ ground state \cite{Bunzli10.01,Bi16.01}.

Therefore, to accurately determine the spacing between the ${}^4I_{15/2}$ and ${}^4F_{9/2}$ energy levels using peak positions, we need to use transitions from the excited states into the same ${}^6H_{15/2}$ sub-level. This however is difficult to accomplish when the transitions to the different sub levels are overlapping from thermal and/or inhomogeneous broadening. This broadening can be minimized either by using low temperatures \cite{Bi16.01} or a host material with a strong crystal field \cite{Bunzli10.01}. Alternatively, by integrating over the full peak (as we do in this study) we can account for all transitions from the ${}^4I_{15/2}$ and ${}^4F_{9/2}$ energy levels into the ${}^6H_{15/2}$ state and obtain an accurate energy difference.

\begin{figure}
 \centering
 \includegraphics{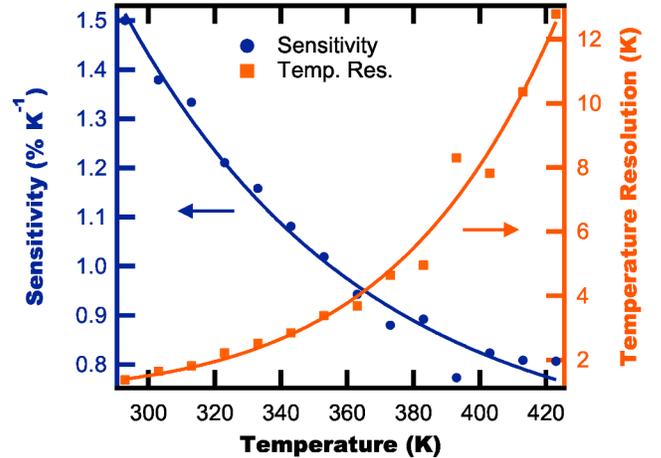}
 \caption{Sensitivity and temperature resolution of our TCT method as a function of temperature.}
 \label{fig:sensres}
\end{figure}

Having determined the temperature response of the intensity ratio, we can now calculate the sensitivity of [Y$_{1-x}$Dy$_x$(acac)$_3$(phen)] and the temperature resolution of our TCT technique. The sensitivity $S$, of a TCT phosphor is given by \cite{Brites12.01}

\begin{align}
S=\frac{100}{r}\frac{\partial r}{\partial T}, \label{eqn:sens}
\end{align}
where $r$ is the intensity ratio and the sensitivity is given in units of \% K$^{-1}$. To calculate the sensitivity we first take the derivative of the intensity ratio function (Equation \ref{eqn:Irat}) and substitute in the average amplitude and energy difference from Table \ref{tab:param}. We then divide the calculated derivative by the experimentally obtained ratio (for the 10 mol\% samples) and compute Equation \ref{eqn:sens} with Figure \ref{fig:sensres} showing the computed sensitivity. From Figure \ref{fig:sensres} we find that the sensitivity has a maximum value of 1.5 \% K$^{-1}$, which is comparable to many other temperature sensing phosphors \cite{Brites12.01}.

In addition to calculating the sensitivity of our phosphor we also consider the temperature resolution $\Delta T$, of our technique, which not only depends on the sensitivity of the phosphor, but also the experimental uncertainty of the optical detection system. The temperature resolution can be defined as the temperature change required to change the ratio by an amount equal to the experimental uncertainty. Mathematically this takes the form of

\begin{align}
\Delta T=\sigma_r\left(\frac{\partial r}{\partial T}\right)^{-1},
\end{align}
where $\sigma_r$ is the experimental uncertainty in the ratio.

Using the calculated derivative  of the ratio and the ratio uncertainty for the 10 mol\% sample we calculate the temperature resolution and show the results in Figure \ref{fig:sensres}. From Figure \ref{fig:sensres} we find that at room temperature the temperature resolution is approximately 1.8 K and that it increases with temperature. This behavior is primarily due to the effects of thermal quenching, which is found to increase the relative uncertainty of the measured ratio as the temperature is increased.

\subsection{Photodegradation}
In the previous section we measured the 455 nm/482 nm intensity ratios of [Y$_{1-x}$Dy$_x$(acac)$_3$(phen)] at different temperatures and determined the calibration parameters for the material's temperature response. While this analysis is the main purpose of our study we also observe significant yellowing of the [Y$_{1-x}$Dy$_x$(acac)$_3$(phen)] powder during extended UV exposure, which suggests that the material undergoes significant photodegradation. 

Given that [Y$_{1-x}$Dy$_x$(acac)$_3$(phen)] consists of either a Dy or Y ion connected to organic ligands, photodegradation will occur solely due to the ligands, implying that the Dy spectral ratios should be unaffected (assuming the ligand stays transparent in the 450 nm -- 500 nm range). However, while the ratios should be unaffected, the overall intensity will be decreased as the ligand acts as an ``antenna'' for the Dy ion \cite{Feng09.01} and damage to the ligand will decrease the efficiency of energy transfer between the ligand and Dy ion.

\begin{figure}
 \centering
 \includegraphics{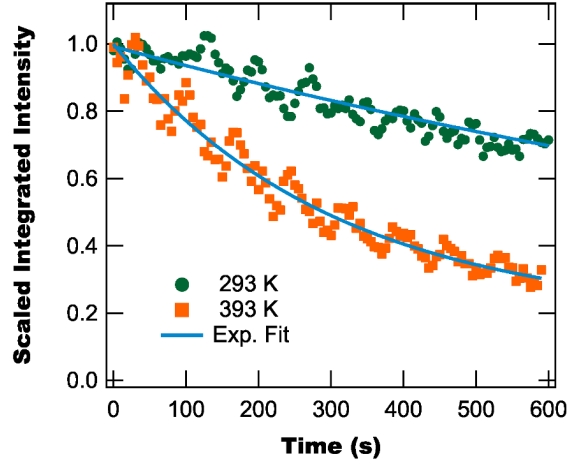}
 \caption{Scaled integrated intensity (465 nm -- 500 nm) as a function of time during 355 nm exposure for two different temperatures. The material is found to photodegrade under 355 nm illumination with the decay rate increasing with temperature. Note that the data is smoothed using a binomial algorithm for clarity.}
 \label{fig:deg}
\end{figure}

To characterize this decrease in energy transfer efficiency (and confirm that the ratios are unaffected) we measure the emission spectra between 450 nm and 500 nm for a 10 mol\% sample as a function of time during constant UV illumination at two different temperatures (293 K and 393 K). For both degradation and excitation we use the same 355 nm pulsed laser with the fluence per pulse being 89 mJ/cm$^2$ (average intensity of 0.891 W/cm$^2$).  Figure \ref{fig:deg} shows the scaled integrated intensity of the 482 nm peak (integration range 465 nm to 500 nm, smoothed for clarity) as a function of time for both 293 K and 393 K, with the intensity scaled such that the initial intensity is 1.

From Figure \ref{fig:deg} we find that the scaled integrated intensity of the 482 nm peak decays exponentially with time, with the higher temperature being found to decay more quickly.  Fitting the decay curves in Figure \ref{fig:deg} to a simple exponential function we find decay rates of  $\gamma=1.40(\pm0.26)\times10^{-3}$ s$^{-1}$  for the 293 K curve and  $\gamma=3.29(\pm0.56)\times10^{-3}$ s$^{-1}$ for the 393 K curve. 

Comparing the decay rates between the two temperatures we find that by increasing the temperature by 100 K the ligand decays 2.35 $\times$ more quickly than at room temperature. This temperature dependence suggests that photodegradation is due to an energy barrier process. To estimate this energy barrier we first assume that the decay rate depends on temperature as an Arrhenius function:

\begin{align}
\gamma=\gamma_0e^{-\Delta E_D/kT},
\end{align}
where $\gamma_0$ is the asymptotic decay rate and $\Delta E_D$ is the photodegradation energy barrier. With this assumption we calculate the ratio of decay rates to be,

\begin{align}
\frac{\gamma_1}{\gamma_2}=\exp\left[-\frac{\Delta E_D}{k}\left(\frac{1}{T_1}-\frac{1}{T_2}\right)\right], \label{eqn:grat}
\end{align}
where $\gamma_1$ is the decay rate at temperature $T_1$ and $\gamma_2$ is the decay rate at temperature $T_2$. Equation \ref{eqn:grat} can  be rearranged to give the energy barrier to be

\begin{align}
\Delta E_D=-k\frac{T_1T_2}{T_2-T_1}\ln \left(\frac{\gamma_1}{\gamma_2}\right).
\end{align}
Substituting in our experimental values we can estimate the energy barrier to be $\Delta E_D=0.084\pm 0.011$ eV. Note that a more accurate determination -- using degradation at many different temperatures -- of this energy barrier is beyond the scope of this current study.

In addition to determining the temperature dependence of (acac)$_3$(phen)'s photodegradation, we also compare its photostability to other organic materials by computing the commonly used figure of merit for organic dyes \cite{Anderson15.04}:

\begin{align}
\text{FoM}&=\frac{B}{\sigma}
\\ &=\frac{\langle I \rangle}{\gamma \hbar\omega},\label{eqn:fom}
\end{align}
where $\sigma$ is the undamaged absorbance cross section, $B$ is the degradation's inverse quantum efficiency, $\langle I\rangle$ is the average pump intensity, $\gamma$ is the decay rate, and $\hbar\omega$ is the photon energy of the degradation source. Using Equation \ref{eqn:fom} and the values from our room temperature (293 K) measurement we determine a FoM of  $1.13(\pm 0.21)\times 10^{21}$ cm$^{-2}$, which is on the same order as a large number of organic materials, but almost two orders of magnitude below the most photostable organic materials \cite{Anderson15.04}.

\begin{figure}
 \centering
 \includegraphics{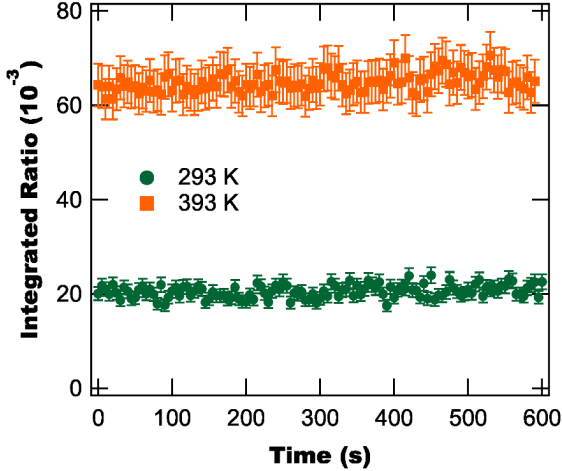}
 \caption{Integrated ratio (455 nm/482 nm) as a function of time during photodegradation.  The ratio is found to be invariant within experimental uncertainty for the entire degradation time period.}
 \label{fig:decrat}
\end{figure}

With the influence of (acac)$_3$(phen)'s photodegradation on the emission intensity now determined, we turn to considering how (acac)$_3$(phen)'s photodegradation affects the intensity ratio between the 455 nm peak and the 482 nm peak. Using the same calculation method described in Section \ref{sec:rat} we calculate the integrated peak ratio as a function of temperature for both temperatures as shown in Figure \ref{fig:decrat}. From Figure \ref{fig:decrat} we find that for both 293 K and 393 K the integrated ratio is invariant within uncertainty for the entire time frame of photodegradation. This result confirms our initial hypothesis that photodegradation won't influence the intensity ratios.

\subsection{Dynamic Heating Test}
Having now determined the temperature dependence of the integrated ratio and shown that the ratio is invariant during ligand photodegradation, we now demonstrate the use of [Y$_{1-x}$Dy$_x$(acac)$_3$(phen)] as a TCT phosphor. To do so we heat a sample (10 mol\%) and measure the luminescence spectrum and TC temperature as a function of time using the gated spectroscopy system. 

\begin{figure}
 \centering
 \includegraphics{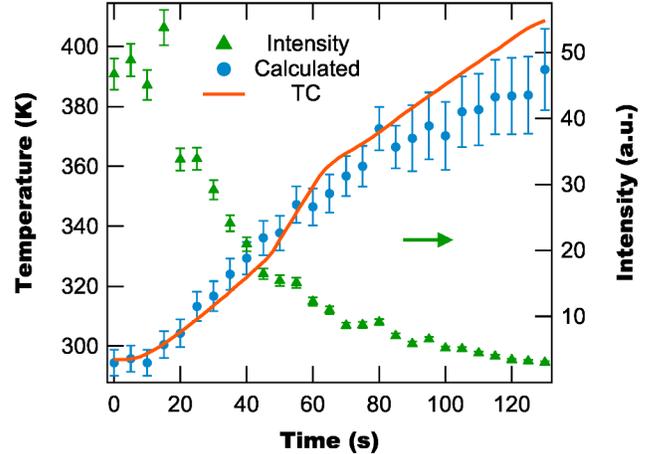}
 \caption{Temperature as a function of time as measured by a TC and calculated using TCT with [Y$_{0.9}$Dy$_{0.1}$(acac)$_3$(phen)]. For comparison we also include the intensity at 482 nm as a function of time. We find that as the phosphor intensity approaches zero the calculated temperature becomes more noisy, with the peak accurate temperature near 370 K.}
 \label{fig:Tcalc}
\end{figure}

After heating we compute the integrated intensity ratio as a function of time and calculate the temperature using Equation \ref{eqn:dist} and the calibration values. Figure \ref{fig:Tcalc} shows the calculated temperature, TC value, and the peak intensity at 482 nm as a function of time. From Figure \ref{fig:Tcalc} we find that the calculated temperature is within uncertainty of the TC temperature for temperatures up to about 370 K after which point the calculated temperatures are more noisy. This noise in temperatures is due to the emission intensity decreasing to the point where the SNR is too small to obtain accurate temperatures.

From Figures \ref{fig:TI} and \ref{fig:Tcalc} it should be noted that the maximum accurate temperature is lower in Figure \ref{fig:Tcalc} than in Figure \ref{fig:TI}. The reason for this difference is related to the different camera settings between the static calibration measurements and the dynamic heating measurements. During the calibration measurements the camera gain is adjusted for each temperature to account for the decrease in intensity, while during dynamic heating the gain is fixed for the entire run. This leads to the intensity quickly falling below the accurate detection limit. We are currently working on developing new software which will dynamically adjust the camera gain allowing for a larger temperature range.

\section{Conclusions}
We develop a TCT phosphor based on a [Y$_{1-x}$Dy$_x$(acac)$_3$(phen)] molecular crystals. The emission (using an excitation wavelength of 355 nm) is found to contain four peaks originating from the Dy ion and a broad background peak arising from the ligand. The emission from the Dy ion at room temperature is found to have a lifetime ranging from 6 $\mu$s to 16 $\mu$s depending on Dy concentration, and the emission from the ligand at room temperature has a lifetime of approximately 600 ns. This difference in lifetime allows us to remove the ligand's contribution from the emission spectra by using time-gated spectroscopy.

Using this technique we characterize the material's emission for four different concentrations and 14 different temperatures. From these measurements we find that the material displays both concentration and thermal quenching, which results in decreased emission intensity and luminescence lifetimes at high concentrations and temperatures. With regards to concentration quenching we find that the peak intensity occurs for a Dy concentration of $\approx 10$ mol\% and that the lifetime as a function of concentration appears to follow the functionality of Dexter electron exchange quenching. While the estimated ion spacing is found to be the correct order of magnitude for the Dexter exchange mechanism, a more in depth study of the concentration dependence is needed to verify this hypothesis.

We also find that the effect of thermal quenching on [Y$_{1-x}$Dy$_x$(acac)$_3$(phen)]'s emission is consistent with results seen in other lanthanide-doped complexes. We find that the emission intensity decays exponentially in the temperature range tested (with the 1/e intensity decay temperature being $36.89 \pm 0.84$ K) and that the emission lifetime obeys Equation \ref{eqn:LTTQ} with a thermal quenching barrier of $\approx 0.2$ eV. Due to this thermal quenching we find that -- for our current spectroscopy system -- we can accurately measure the emission to a temperature of $\approx$423 K. 

Additionally, we find that the organic ligand photodegrades under UV illumination resulting in a decrease of the Dy emission peaks as the ligand acts as an ``antenna'' for the Dy ion. Despite this decrease in luminescence efficiency, due to photodegradation, we find that the ratio between the 455 nm and 482 nm peaks is unchanged.  

With the emission of [Y$_{1-x}$Dy$_x$(acac)$_3$(phen)] characterized, we next demonstrate its use as a TCT phosphor by heating a 10 mol\% sample dynamically and measuring the emission as a function of time using our time-gated spectroscopy system.  Based on the emission spectra (and using the known calibration parameters) we compute the temperature and compare the calculated value to that measured by a thermocouple with both temperatures found to be within uncertainty of each other up to a temperature of approximately 370 K, with the main limitation due to the current spectroscopy systems software. To remove this limitation we are currently developing improved software, which will allow for a higher temperature range with our initial target maximum temperature being 423 K.

In summary, we find that [Y$_{1-x}$Dy$_x$(acac)$_3$(phen)] forms organic molecular crystals with appropriate spectroscopic properties for use as a TCT phosphor. Our next step -- towards using [Y$_{1-x}$Dy$_x$(acac)$_3$(phen)] as a TCT phosphor during shock compression of heterogeneous materials -- is to perform fast laser heating of the phosphor in order to demonstrate its ability to measure temperature during short time scales. At the same time we are also testing other Dy$^{3+}$-doped Yttrium ternary complexes to determine how their thermal performance compares to [Y$_{1-x}$Dy$_x$(acac)$_3$(phen)].

\section*{Acknowledgements}
This work was supported by the Air Force Office of Scientific Research, Award \# FA9550-15-1-0309 to Washington State University.


\begin{thebibliography}{10}
\newcommand{\enquote}[1]{``#1''}

\bibitem{Chen14.01}
M.~Chen, S.~You, K.~Suslick, and D.~Dlott, \enquote{Hot spots in energetic
  materials generated by infared and ultrasound, detected by thermal imaging
  microscopy,} Rev. Sci. Instrum \textbf{85}, 023705 (2014).

\bibitem{Chen14.02}
M.~Chen, S.~You, K.~Suslick, and D.~Dlott, \enquote{Hot spot generation in
  energetic materials created by long-wavelength infared radiation,} App. Phys.
  Lett. \textbf{104}, 061907 (2014).

\bibitem{Pokharel14.01}
R.~Pokharel, J.~Lind, A.~Kanjarla, R.~Lebensohn, S.~Li, P.~Kenesei, R.~Suter,
  and A.~Rollett, \enquote{Polycrystal plasticity: comparison between
  grain-scale observations of deformation and simulations,} Annual Review of
  Condensed Matter Physics \textbf{5}, 317--346 (2014).

\bibitem{An13.01}
Q.~An, W.~Goddard, S.~Zybin, A.~Jaramillo-Botero, and T.~Zhou, \enquote{Highly
  shocked polymer bonded explosiuves at a nonplanar interface: hot-spot
  formation leading to detonation,} Journal of Physical Chemistry C
  \textbf{117}, 26551--26561 (2013).

\bibitem{Wu11.01}
Y.~Wu and F.~Huang, \enquote{A microscopic model for predicting hot-spot
  ignition of granular energetic crystals in response to drop-weight impacts,}
  Mechanics of Materials \textbf{43}, 835--852 (2011).

\bibitem{Rae02.01}
P.~Rae, H.~Goldrein, S.~Palmer, J.~Field, and A.~Lewis, \enquote{Quasi-static
  studies of the deformation and failure of beta-hmx based polymer bonded
  explosives,} Proceedings of the Royal Society a-Mathematical Physical and
  Engineering Sciences \textbf{458}, 743--762 (2002).

\bibitem{Gruzdkov01.01}
Y.~Gruzdkov and Y.~Gupta, \enquote{Vibrational properties and structure of
  pentaerythritol tetranitrate,} J. Phys. Chem. A \textbf{105}, 6197--6202
  (2001).

\bibitem{Gupta95.01}
Y.~Gupta, \enquote{Recent developments to understand molecular-changes in
  schocked energetic materials,} Journal De Physique IV \textbf{5}, 345--356
  (1995).

\bibitem{Mellor93.01}
A.~Mellor, D.~Wiegand, and K.~Isom, \enquote{Hot-spot histories in energetic
  materials,} Structure and Properties of Energetic Materials \textbf{296},
  293--298 (1993).

\bibitem{Field92.01}
J.~Field, \enquote{Hot-spot ignition mechanisms for explosives,} Acc. Chem.
  Res. \textbf{25}, 489--496 (1992).

\bibitem{Field92.02}
J.~Field, N.~Bourne, S.~Palmer, S.~Walley, and J.~Smallwood, \enquote{Hot-spot
  ignition mechanisms for explosives and propellants,} Philosophical
  Transactions of the Royal Society of London-Mathematical Physical and
  Engineering Sciences \textbf{339}, 269--283 (1992).

\bibitem{Field82.01}
J.~Field, G.~Swallowe, and S.~Heavens, \enquote{Ignition mechanisms of
  explosives during mechanical deformation,} Proceedings of the Royal Society
  of London Series A \textbf{382}, 231 (1982).

\bibitem{Winter75.01}
R.~Winter and E.~Faber, \enquote{The role of localized plastic flow in the
  impact initiation of explosives,} Proceedings of the Royal Society of London
  Series A \textbf{343}, 399--413 (1975).

\bibitem{Chaudhri74.01}
M.~Chaudhri and J.~Field, \enquote{The role of rapidly compressed gas pockets
  in the initiation of condensed explosives,} Proceedings of the Royal Society
  of London \textbf{340}, 113--128 (1974).

\bibitem{Luo12.01}
S.~N. {Luo}, B.~J. {Jensen}, D.~E. {Hooks}, K.~J. {Ramos}, J.~D. {Yeager},
  K.~{Kwiatkowski}, T.~{Shimada}, D.~A. {Fredenburg}, and K.~{Fezzaa},
  \enquote{{Ultrafast, high resolution, phase contrast imaging of shock
  response with synchrotron radiation: opportunities and challenges},} in
  \enquote{APS Meeting Abstracts,}  (2012).

\bibitem{Jensen14.01}
B.~J. Jensen, K.~J. Ramos, A.~J. Iverson, J.~Bernier, C.~A. Carlson, J.~D.
  Yeager, K.~Fezzaa, and D.~E. Hooks, \enquote{Dynamic experiment using impulse
  at the advanced photon source,} Journal of Physics: Conference Series
  \textbf{500}, 042001 (2014).

\bibitem{Jensen12.01}
B.~J. Jensen, S.~N. Luo, D.~E. Hooks, K.~Fezzaa, K.~J. Ramos, J.~D. Yeager,
  K.~Kwiatkowski, T.~Shimada, and D.~M. Dattelbaum, \enquote{Ultrafast, high
  resolution, phase contrast imaging of impact response with synchrotron
  radiation,} AIP Advances \textbf{2}, 012170 (2012).

\bibitem{Yeager12.01}
J.~Yeager, S.~Luo, B.~Jensen, K.~Fezzaa, D.~Montgomery, and D.~Hooks,
  \enquote{High-speed synchrotron x-ray phase contrast imaging for analysis of
  low-z composite microstructure,} Composites Part A: Applied Science and
  Manufacturing \textbf{43}, 885 -- 892 (2012).

\bibitem{Luo12.02}
S.~N. Luo, B.~J. Jensen, D.~E. Hooks, K.~Fezzaa, K.~J. Ramos, J.~D. Yeager,
  K.~Kwiatkowski, and T.~Shimada, \enquote{Gas gun shock experiments with
  single-pulse x-ray phase contrast imaging and diffraction at the advanced
  photon source,} Review of Scientific Instruments \textbf{83}, 073903 (2012).

\bibitem{DCSweb}
\enquote{http://www.dcs-aps.wsu.edu/,} .

\bibitem{DCS12.01}
\enquote{New research opportunities in dynamic compression science,} Tech.
  rep., Institute for Shock Physics, Washington State University (2012).

\bibitem{Juboori13.01}
A.~Z.~M. Al-Juboori, \enquote{Rare earth (Sm$^{3+}$ and Dy$^{3+}$)-doped
  gadolinium oxide nanomaterials for luminescence thermometry,} Physica Scripta
   (2013).

\bibitem{Heyes06.01}
A.~Heyes, S.~Seefeldt, and J.~Feist, \enquote{Two-colour phosphor thermometry
  for surface temperature measurement,} Optics \& Laser Technology \textbf{38},
  257--265 (2006).

\bibitem{Khalid08.01}
A.~Khalid and K.~Kontis, \enquote{Thermographic phosphors for high temperature
  measurements: Principles, current state of the art and recent applications,}
  Sensors \textbf{8}, 5673--5744 (2008).

\bibitem{Kontis07.01}
K.~Kontis, \enquote{A review of some current research on pressure sensitive and
  thermographic phosphor techniques,} The Aeronautical Journal \textbf{3162},
  495--508 (2007).

\bibitem{Allison97.01}
S.W., Allison, and G.~Gillies, \enquote{Remote thermometry with thermographic
  phosphors: Instrumentation and applications,} Review Scientific Instruments
  \textbf{68}, 2615--2650 (1997).

\bibitem{Cates03.01}
M.~Cates, S.~Allison, S.~Jaiswal, and D.~Beshears, \enquote{YAG:Dy and YAG:Tm
  fluorescence to 1700 $^\circ$c,} in \enquote{The 49th International
  Instrumentation Symposium -- The Instrumentation, Systems, and Automation
  Society,}  (Orlando, FL, 2003).

\bibitem{Gross89.01}
G.~Gross, A.~Smith, and M.~Post, \enquote{Surface thermometry by laser induced
  fluorescence,} Review Scientific Instruments \textbf{60}, 12 (1989).

\bibitem{Kontis02.01}
K.~Kontis, Y.~Syogenji, and N.~Yoshikawa, \enquote{Surface thermometry by laser
  induced fluorescence of Dy$^{3+}$ :YAG,} The Aeronautical Journal
  \textbf{106}, 453--457 (2002).

\bibitem{Kontis03.01}
K.~Kontis, \enquote{Surface heat transfer measurements inside a supersonic
  combustor by laser induced fluorescence,} Journal of Thermophysics and Heat
  Transfer \textbf{17}, 320--325 (2003).

\bibitem{Hasegawa07.01}
R.~Hasegawa, I.~Sakata, H.~Yanagihara, B.~Johansson, A.~Omrane, and M.~Alden,
  \enquote{Two-dimensional gas phase temperature measurements using phosphor
  thermometry,} App. Phys. B \textbf{88}, 291--296 (2007).

\bibitem{Sarner05.01}
G.Sarner, A.~Omrane, H.~Seyfried, M.~Richter, H.~Schmidt, and M.~Alden,
  \enquote{Laser diagnostics applied to a full-size fighter jet afterburner,}
  in \enquote{European Combustion Meeting,}  (2005).

\bibitem{Feist02.01}
J.~Feist, A.~Heyes, and S.~Seefeldt, \enquote{Thermographic phosphors for gas
  turbines: instrumentation development and measurement uncertainties,} in
  \enquote{11th International symposium on Application of Laser Techniques to
  Fluid Mechanics,}  (Lisbon, Portugal, 2002).

\bibitem{Feist01.01}
J.~P. Feist, A.~L. Heyes, and J.~R. Nicholls, \enquote{Phosphor thermometry in
  an electron beam physical vapour deposition produced thermal barrier coating
  doped with dysprosium,} in \enquote{Proceedings of the Institution of
  Mechanical Engineers,} , vol. 215 (2001), vol. 215, pp. 333--341.

\bibitem{Kumar15.01}
J.~P. Kumar, G.~Ramgopal, Y.~Vidya, K.~Anantharaju, B.~D. Prasad, S.~Sharma,
  S.~Prashantha, H.~Nagaswarupa, D.~Kavyashree, and H.~Nagabhushana,
  \enquote{Green synthesis of Y$_2$O$_3$:Dy$^{3+}$ nanophosphor with enhanced
  photocatalytic activity,} Spectrochimica Acta Part A: Molecular and
  Biomolecular Spectroscopy \textbf{149}, 687 -- 697 (2015).

\bibitem{Gupta13.01}
S.~K. Gupta, M.~Mohapatra, V.~Natarajan, and S.~V. Godbole,
  \enquote{Photoluminescence investigations of the near white light emitting
  perovskite ceramic SrZrO$_3$:Dy$^{3+}$ prepared via gel-combustion route,}
  International Journal of Applied Ceramic Technology \textbf{10}, 593--602
  (2013).

\bibitem{Jayasimhadri11.01}
M.~Jayasimhadri, B.~V. Ratnam, K.~Jang, H.~S. Lee, B.~Chen, S.-S. Yi, J.-H.
  Jeong, and L.~Rama~Moorthy, \enquote{Combustion synthesis and luminescent
  properties of nano and submicrometer-size Gd$_2$O$_3$:Dy$^{3+}$ phosphors for
  white leds,} International Journal of Applied Ceramic Technology \textbf{8},
  709--717 (2011).

\bibitem{Ratnam10.01}
B.~Ratnam, M.~Jayasimhadri, K.~Jang, H.~Sueb~Lee, S.-S. Yi, and J.-H. Jeong,
  \enquote{White light emission from NaCaPO$_4$:Dy$^{3+}$ phosphor for
  ultraviolet-based white light-emitting diodes,} Journal of the American
  Ceramic Society \textbf{93}, 3857--3861 (2010).

\bibitem{Kaur14.01}
J.~Kaur, Y.~Parganiha, V.~Dubey, D.~Singh, and D.~Chandrakar,
  \enquote{Synthesis, characterization and luminescence behavior of
  ZrO$_2$:Eu$^{3+}$, Dy$^{3+}$ with variable concentration of Eu and Dy doped
  phosphor,} Superlattices and Microstructures \textbf{73}, 38 -- 53 (2014).

\bibitem{Akita15.01}
Y.~Akita, T.~Harada, R.~Sasai, K.~Tomita, and K.~Nishiyama, \enquote{Emission
  properties of Ln (Eu, Tb, Dy, Er)-doped \{Y$_2$O$_3$\} nanoparticles
  synthesized by surfactant-assembly and their applications in visible
  color-tuning,} Journal of Photochemistry and Photobiology A: Chemistry
  \textbf{299}, 87 -- 93 (2015).

\bibitem{Anderson16.07}
B.R.~Anderson, R.~Gunawidjaja, P.~Price, and H.~Eilers, \enquote{Spectroscopic determination of thermal impulse in sub-second heating events using lanthanide-doped oxide precursors and phenomenological modeling,} Journal of Applied Physics, \textbf{120}, 083102 (2016).

\bibitem{Gunawidjaja16.01}
R.~Gunawidjaja, B.R.~Anderson, H.~D. y~Riega, and H.~Eilers, \enquote{Sub-second
  laser heating of thermal impulse sensors,} AIP Conf. Proc. p. pending
  publication (2016).

\bibitem{Carnall68.01}
R.~K. Carnall~W.T., \enquote{Electronic energy levels in the trivalent
  lanthanide aquo ions,} J. Chem Phys. \textbf{49}, 4424--4442 (1968).

\bibitem{Carnall79.01}
W.~Carnall, \enquote{The absorption and fluorescence spectra of rare earth ions
  in solution,} in \enquote{Handbook on the physics and chemistry of rare
  earths,} , K.~Gschneidner and L.~Eyring, eds. (Elsevier BV, Amsterdam, 1979),
  chap.~24.

\bibitem{Bunzli10.01}
J.-C.~G. B{\"u}nzli and S.~V. Eliseeva, \enquote{Basics of lathanide
  photophysics,} in \enquote{Lanthanide Luminescence: Photophysical, Analytical
  and Biological Aspects,} , P.~Hanninen and H.~Harma, eds. (Springer-Verlag,
  2010).

\bibitem{Bunzli10.02}
J.~B{\"u}nzli, \enquote{Lanthanide luminescence for biomedical analyses and
  imaging,} Chem. Rev. \textbf{110}, 2729--2755 (2010).

\bibitem{Sager65.01}
N.~F. W.~F.~Sager and F.~A. Serafin, \enquote{Substituent effects on
  intramolecular energy transfer. I. absorption and phosphorescence spectra of
  rare earth $\beta$-diketone chelates,} J Phys. Chem \textbf{69}, 1092 (1965).

\bibitem{Dexter53.01}
D.~Dexter, \enquote{A theory of sensitized luminescence in solids,} J. Chem.
  Phys. \textbf{21}, 836 (1953).

\bibitem{Xin04.01}
H.~Xin, M.~Shi, X.~C. Gao, Y.~Y. Huang, Z.~L. Gong, D.~B. Nie, H.~Cao, Z.~Q.
  Bian, F.~Y. Li, and C.~H. Huang, \enquote{The effect of different neutral
  ligands on photoluminescence and electroluminescence properties of ternary
  terbium complexes,} The Journal of Physical Chemistry B \textbf{108},
  10796--10800 (2004).

\bibitem{Klink00.01}
S.~I. Klink, \enquote{Synthesis and photophysics of light-converting lanthanide
  complexes,} Ph.D. thesis, University of Twente (2000).

\bibitem{Latva97.01}
M.~Latva, H.~Takalo, V.-M. Mukkala, C.~Matachescu, J.~C. Rodriguez-Ubis, and
  J.~Kankare, \enquote{Correlation between the lowest triplet state energy
  level of the ligand and lanthanide(III) luminescence quantum yield,} Journal
  of Luminescence \textbf{75}, 149 -- 169 (1997).

\bibitem{Feng09.01}
J.~Feng, L.~Zhou, S.-Y. Song, Z.-F. Li, W.-Q. Fan, L.-N. Sun, Y.-N. Yu, and
  H.-J. Zhang, \enquote{A study on the near-infrared luminescent properties of
  xerogel materials doped with dysprosium complexes,} Dalton Trans. pp.
  6593--6598 (2009).

\bibitem{Adronov00.01}
A.~Adronov and J.~M. Frechet, \enquote{Light-harvesting dendrimers,} Chem Comm
  pp. 1701--1710 (2000).

\bibitem{Dickens98.01}
R.~S. Dickens, J.~A.~K. Howard, C.~L. Maupin, J.~M. Moloney, D.~Parker, R.~D.
  Peacock, J.~P. Riehl, and G.~Siligardi, \enquote{Ground and excited state
  chiroptical properties of enantiopure macrocyclic tetranaphthyl lanthanide
  complexes: controlled modulation of the frequency and polarisation of emitted
  light,} New J. Chem. \textbf{22}, 891--899 (1998).

\bibitem{Bunzli89.01}
J.~B{\"u}nzli, \enquote{Luminescent probes,} in \enquote{Lanthanide Probes in
  Life Chemical and Earth Sciences: Theory and Practice,} , J.~Bunzli and
  G.~Choppin, eds. (Elsevier, 1989), pp. 219--293.

\bibitem{Shavaleev10.01}
N.~M. Shavaleev, S.~V. Eliseeva, R.~Scopelliti, and J.-C.~G. B{\"u}nzli,
  \enquote{N-aryl chromophore ligands for bright europium luminescence,}
  Inorganic Chemistry \textbf{49}, 3927--3936 (2010). PMID: 20302272.

\bibitem{Eliseeva10.01}
S.~V. Eliseeva and J.-C.~G. B{\"u}nzli, \enquote{Lanthanide luminescence for
  functional materials and bio-sciences,} Chem. Soc. Rev. \textbf{39}, 189--227
  (2010).

\bibitem{Puntus09.01}
L.~N. Puntus, K.~A. Lyssenko, I.~S. Pekareva, and J.-C.~G. B{\"u}nzli,
  \enquote{Intermolecular interactions as actors in energy-transfer processes
  in lanthanide complexes with 2,2'-bipyridine,} The Journal of Physical
  Chemistry B \textbf{113}, 9265--9277 (2009). PMID: 19522489.

\bibitem{Bunzli14.01}
J.-C.~G. B{\"u}nzli, \enquote{Review: Lanthanide coordination chemistry: from
  old concepts to coordination polymers,} Journal of Coordination Chemistry
  \textbf{67}, 3706--3733 (2014).

\bibitem{Bunzli16.01}
J.-C.~G. B{\"u}nzli, \enquote{Lanthanide light for biology and medical
  diagnosis,} Journal of Luminescence \textbf{170, Part 3}, 866 -- 878 (2016).
  Light, Energy and Life.

\bibitem{Bunzli15.01}
J.-C.~G. B{\"u}nzli, \enquote{On the design of highly luminescent lanthanide
  complexes,} Coordination Chemistry Reviews \textbf{293--294}, 19 -- 47 (2015).
  41st International Conference on Coordination Chemistry, Singapore, July
  2014.

\bibitem{Forster65.01}
T.~Forster, \enquote{Delocalized excitation and excitation transfer,} in
  \enquote{Modern Quantum Chemistry. Istanbul Lectures. Part III: Action of
  Light and Organic Crystals,} , vol.~3 (Academic Press, 1965), pp. 93--137.

\bibitem{Forster59.01}
T.~Forster, \enquote{10th spiers memorial lecture. transfer mechanisms of
  electronic excitation,} Discussions of the Faraday Society \textbf{27}, 7--17
  (1959).

\bibitem{Forster48.01}
T.~Forster, \enquote{Zwischenmolekulare energiewanderung und fluoreszenz,} Ann.
  Phys. \textbf{2}, 55 (1948).

\bibitem{Forster46.01}
T.~Forster, \enquote{Energiewanderung und fluoreszenz,} Naturwissenschaften
  \textbf{33}, 166--175 (1946).

\bibitem{Dexter54.01}
D.~Dexter and J.~H. Schulman, \enquote{Theory of concentration quenching in
  inorganic phosphors,} J . Chem. Phys. \textbf{22}, 1063 (1954).

\bibitem{Dexter51.01}
D.~Dexter and W.~Heller, \enquote{Capture and collision processes for excitons
  in alkali halides,} Phys. Rev. \textbf{84}, 377 (1651).

\bibitem{Heller51.01}
W.~R. Heller and A.~Marcus, \enquote{A note on the propagation of excitation in
  an idealized crystal,} Phys. Rev. \textbf{84}, 809 (1951).

\bibitem{Sato70.01}
S.~Sato and M.~Wada, \enquote{Relations between intramolecular energy transfer
  efficiencies and triplet state energies in rare earth $\beta$-diketone
  chelates,} Bulletin of the Chemical Society of Japan \textbf{43}, 1955--1962
  (1970).

\bibitem{Tang15.01}
J.~Tang and P.~Zhang, \emph{Lanthanide Single Molecule Magnets} (Springer,
  2015).

\bibitem{Chen11.02}
G.-J. Chen, C.-Y. Gao, J.-L. Tian, J.~Tang, W.~Gu, X.~Liu, S.-P. Yan, D.-Z.
  Liao, and P.~Cheng, \enquote{Coordination-perturbed single-molecule magnet
  behaviour of mononuclear dysprosium complexes,} Dalton Trans. \textbf{40},
  5579 (2011).

\bibitem{Choy07.01}
W.~C. Choy and D.~Tao, \enquote{Rare-earth complexes and its applications in
  organic light emitting diodes,} in \enquote{Solid State Chemistry Research
  Trends,} , R.~W. Buckley, ed. (Nova Science Publishers Inc, 2007), chap.~2.

\bibitem{Brito09.01}
H.~Brito, O.~Malta, M.~Felinto, and E.~Teotonio, \enquote{Luminescence
  phenomena involving metal enolates,} in \enquote{The Chemistry of Metal
  Enolates,} , J.~Zabicky, ed. (Wiley, 2009), The Chemistry of Functional
  Groups, chap.~3.

\bibitem{Peng05.01}
C.~Y. Peng, H.~J. Zhang, J.~B. Yu, Q.~G. Meng, L.~S. Fu, H.~R. Li, L.~N. Sun,
  and X.~M. Guo, \enquote{Synthesis, characterization, and luminescence
  properties of the ternary europium complex covalently bonded to mesoporous
  sba-15,} J Phys. Chem B \textbf{109}, 15278 (2005).

\bibitem{Xin06.01}
H.~Xin, Y.~Ebina, R.~Ma, K.~Takada, and T.~Sasaki, J Phys. Chem B \textbf{110},
  9863 (2006).

\bibitem{Kharabe15.01}
V.~R. Kharabe, A.~H. Oza, and S.~J. Dhoble, \enquote{Synthesis, PL
  characterizations and concentration quenching effect in Dy$^{3+}$ and
  Eu$^{3+}$ activated LiCaBO$_3$ phosphor,} Luminescence \textbf{30}, 432--438
  (2015).

\bibitem{Joos12.01}
J.~J. Joos, K.~W. Meert, A.~B. Parmentier, D.~Poelman, and P.~F. Smet,
  \enquote{Thermal quenching and luminescence lifetime of saturated green
  Sr$_{1-x}$Eu$_x$ Ga$_2$S$_4$ phosphors,} Optical Materials \textbf{34},
  1902--1907 (2012).

\bibitem{Meza14.01}
O.~Meza, E.~Villabona-Leal, L.~Diaz-Torres, H.~Desirena, J.~Rodriguez-Lopez,
  and E.~Perez, \enquote{Luminescence concentration quenching mechanism in
  Gd$_2$O$_3$:Eu$^{3+}$,} J Phys. Chem A \textbf{118}, 1390--1396 (2014).

\bibitem{Ju13.01}
G.~Ju, Y.~Hu, L.~Chen, X.~Wang, and Z.~Mu, \enquote{Concentration quenching of
  persistent luminescence,} Physica B: Condensed Matter \textbf{415}, 1 -- 4
  (2013).

\bibitem{lakow06.01}
J.~R. Lakowicz, \emph{Principles of Fluorescence Spectroscopy} (Springer,
  2006).

\bibitem{Bierwagen16.01}
J.~Bierwagen, S.~Yoon, N.~Gartmann, B.~Walfort, and H.~Hagemann,
  \enquote{Thermal and concentration dependent energy transfer of Eu$^{2+}$ in
  SrAl$_2$O$_4$,} Optical Materials Express \textbf{6} (2016).

\bibitem{Ueda15.01}
J.~Ueda, P.~Dorenbos, A.~J. Bos, A.~Meijerink, and S.~Tanabe, \enquote{Insight
  into the thermal quenching mechanism for Y$_3$Al$_5$O$_12$:Ce$^{3+}$ through
  thermoluminescence excitation spectroscopy,} J. Phys. Chem C \textbf{119},
  25003--25008 (2015).

\bibitem{Yu14.01}
R.~Yu, D.~S. Shin, K.~Jang, Y.~Guo, H.~M. Noh, B.~K. Moon, B.~C. Choi, J.~H.
  Jeong, and S.~S. Yi, \enquote{Luminescence and thermal-quenching properties
  of Dy$^{3+}$-doped Ba$_2$CaWO$_6$ phosphors,} Spectrochimica Acta A
  \textbf{125}, 458--462 (2014).

\bibitem{Feist00.01}
J.~Feist and A.~Heyes, \enquote{The characterization of Y$_2$O$_2$S:Sm powder
  as a thermographic phosphor for high temperature applications,} Measure
  Science Technology \textbf{11}, 942--947 (2000).

\bibitem{Bi16.01}
Y.~Bi, C.~Chen, Y.-F. Zhao, Y.-Q. Zhang, S.-D. Jiang, B.-W. Wang, J.-B. Han,
  J.-L. Sun, Z.-Q. Bian, Z.-M. Wang, and S.~Gao, \enquote{Thermostability and
  photoluminescence of Dy( III ) single-molecule magnets under a magnetic
  field,} Chem. Sci. \textbf{7}, 5020 (2016).

\bibitem{Brites12.01}
C.~D.~S. Brites, P.~P. Lima, N.~J.~O. Silva, A.~Millan, V.~S. Amaral,
  F.~Palacio, and L.~D. Carlos, \enquote{Thermometry at the nanoscale,}
  Nanoscale \textbf{4}, 4799 (2012).

\bibitem{Anderson15.04}
B.~R. Anderson, S.-T. Hung, and M.~G. Kuzyk, \enquote{Wavelength dependence of
  reversible photodegradation of disperse orange 11 dye-doped pmma thin films.}
  J. Opt. Soc. Am. B \textbf{32}, 1043--1049 (2015).

\end{thebibliography}

\end{document}